\documentclass[preprint,nofootinbib,amsmath,prd,aps,superscriptaddress,tightenlines,12pt]{revtex4}
\usepackage{graphicx}
\usepackage{graphics}
\usepackage{amsmath}
\usepackage{hyperref}
\usepackage{bm}

\begin{document}
\setlength\baselineskip{17pt}

\begin{flushright}
\vbox{
\begin{tabular}{l}
ANL-HEP-PR-13-28
\end{tabular}
}
\end{flushright}
\vspace{0.1cm}


\title{\bf Higgs boson decays to quarkonia and the $H\bar{c}c$ coupling}

\vspace*{1cm}

\author{Geoffrey Bodwin}
\email[]{gtb@anl.gov}
\affiliation{High Energy Physics Division, Argonne National Laboratory, Argonne, IL 60439, USA} 
\author{Frank Petriello}
\email[]{f-petriello@northwestern.edu}
\affiliation{High Energy Physics Division, Argonne National Laboratory, Argonne, IL 60439, USA} 
\affiliation{Department of Physics \& Astronomy, Northwestern University, Evanston, IL 60208, USA}
\author{Stoyan Stoynev}
\email[]{Stoyan.Stoynev@cern.ch}
\affiliation{Department of Physics \& Astronomy, Northwestern University, Evanston, IL 60208, USA}
\author{Mayda Velasco}
\email[]{Mayda.Velasco@cern.ch}
\affiliation{Department of Physics \& Astronomy, Northwestern University, Evanston, IL 60208, USA}


  \vspace*{0.3cm}

\begin{abstract}
  \vspace{0.5cm}
  
In this paper we discuss decays of the Higgs boson to quarkonia in
association with a photon.  We identify a new mechanism for producing
such final states in Higgs decays that leads to predictions for
the decay rates that differ by an order of magnitude from previous
estimates. Although the branching ratios for these processes are still
small, the processes are experimentally clean, and the $H \to J/\psi
\, \gamma$ decay should be observable at a 14 TeV LHC. We point out
that quantum interference between two different production mechanisms
makes the decay rates sensitive to the $H\bar{Q}Q$ couplings.
Consequently, measurements of the $H \to J/\psi \, \gamma$ decay rate
would allow one to probe the Higgs-charm coupling directly at the LHC.
We discuss the experimental prospects for the observation of these decays and
for the direct measurement of the $H\bar{c}c$ coupling.
    
\end{abstract}

\maketitle


\section{Introduction\label{sec:intro}}

With the discovery of a new spin-zero boson by the ATLAS and CMS
collaborations at the LHC now firmly
established~\cite{:2012gk,:2012gu,atlas,Chatrchyan:2012jja}, attention
has shifted to understanding the couplings of this particle in order
to determine whether it is the Standard Model (SM) Higgs boson.  Current
studies indicate no significant deviation from SM predictions in any
measured channel~\cite{atlascoup,cmscoup}, and extracted values of the
couplings in well-measured modes, such as $\gamma\gamma$, $WW$ and $ZZ$,
have errors approaching 10--20\%.

In addition to improving the measurements of these channels, the future
LHC program will study rare and difficult-to-reconstruct decay modes.
One example is the decay $H \to \gamma f \bar{f}$, where $f$ denotes any
SM fermion.  This final state can be produced via $H \to \gamma
(\gamma^{*},Z)$ followed by the decay
$(\gamma^{*},Z) \to \bar{f}f$.  If the mass of the final-state fermion
is large enough, then there is also a significant contribution from
the process in which the $H$ couples directly to $\bar{f}f$ and one of
the fermions emits a photon.  When the final-state fermion is either an
electron or muon, then the decay  $H \to \gamma l^+ l^-$, although
rare, offers a very clean experimental signature.  The observability may
be enhanced further by the resonant production processes  $H \to
\gamma V$, where $V$ denotes a vector meson, such as the $J/ \psi$ or
the $\Upsilon(1S)$, with the subsequent decay $V \to l^+ l^-$. These
channels are promising experimentally: a high-$p_T$ photon that
is back-to-back to a di-lepton pair that reconstructs to a resonance
is simple to distinguish from background.
  
In this note, we study the exclusive decays $H \to V\gamma$,
where $V=J/\psi$ or~$\Upsilon(1S)$.  We distinguish two separate
production mechanisms for the quarkonium state:
\begin{itemize} 

\item {\it direct production}, which proceeds through the $h\bar{Q}Q$
coupling, where Q denotes either the charm quark (in the case of the
$J/\psi$) or bottom quark (in the case of the $\Upsilon(1S)$);

\item {\it indirect production}, which proceeds through $H \to \gamma
\gamma^{*}$, with the subsequent transition $\gamma^{*} \to V$.

\end{itemize}
The possibility of direct quarkonium production in Higgs decays was
first pointed out in Ref.~\cite{Keung:1983ac}.  However, to our
knowledge the indirect mechanism has not been studied
previously.\footnote{This production mechanism has only been mentioned
in previous works~\cite{Jia:2008ep}.}  We find that, in the case of
the $J/\psi$, the indirect mechanism leads to SM decay rates that are
much larger than the previously estimated direct decay rate. The most
promising mode for LHC observation is $H \to J/\psi \,\gamma$,
followed by the decay $J/\psi \to \mu^+ \mu^-$.  This mode should be
evident as a clear peak above the continuum background in a 14~TeV,
high-luminosity LHC run. Interestingly, the quantum interference with
the larger indirect amplitude enhances the effect of the
direct-production amplitude and potentially allows the
$H\bar{c}c$ coupling to be constrained directly by measurement of
the branching ratio for $H \to J/\psi \, \gamma$. The $H\bar cc$
coupling is otherwise very difficult to access directly at the LHC. 
In the SM, the interference effect leads to a shift of approximately
30\% in the branching ratio, which is potentially observable
experimentally.  Deviations from SM predictions for the $H\bar{c}c$
coupling can lead to larger shifts and can be either observed or
constrained by a measurement of the $H \to J/\psi \, \gamma$ branching
ratio. A determination of the $H\bar{c}c$ coupling would test whether
the observed Higgs boson couples to second-generation quarks with the
strength that is predicted in the SM. It had been expected that
only third-generation quark couplings would be accessible to
measurements at the LHC.  Since the decay mode $H \to J/\psi \,
\gamma$ can only be accessed with high statistics, the
possibility of using it to measure the $H\bar{c}c$ coupling motivates a
high-luminosity run of the LHC.  We note that the $H \to V\gamma$ modes will 
also fulfill an important role at future high-luminosity $e^+ e^-$ machines.  
Measurements of the $H\bar{c}c$ and $H\bar{b}b$ couplings via the direct
decays $H \to \bar{c}c,\bar{b}b$ leave the overall signs of the
couplings undetermined.  This ambiguity is resolved by the interference
that is present in $H \to V\gamma$, providing us with important
additional information about the properties of the Higgs.\footnote{We
thank Heather Logan for pointing this out to us.}

This paper is organized as follows.  We present in detail the
calculation of the direct- and indirect-production amplitudes in
Sec.~\ref{sec:calc}.  We pay careful attention to the theoretical
uncertainties that enter the prediction.  In particular, we point
out that the indirect amplitude can be calculated very accurately
within the SM and has theoretical uncertainties at the
few-percent level, allowing deviations that are due to the direct-production 
amplitude to be observed reliably.  We present numerical
results and study the effect on the $H \to J/\psi \gamma$ and $H \to
\Upsilon(1S)\,\gamma$ branching ratios of deviations of the
$H\bar{c}c$ and $H\bar{b}b$ couplings from SM values.  In
Sec.~\ref{sec:exp}, we study the experimental prospects for
observation of the $J/\psi\,\gamma$ mode at the LHC.  Careful
estimates of the acceptances and sensitivities are performed for the
case of the 14~TeV LHC. Finally, we summarize our conclusions
in Sec.~\ref{sec:conc}.

\section{Calculation and numerical results}
\label{sec:calc}

In this section, we calculate the rate for the exclusive decays of a
Higgs to a quarkonium and a photon. We include contributions from both
the direct and indirect mechanisms, which are described in
Sec.~\ref{sec:intro}, including their quantum-mechanical
interference.  An accounting of the theoretical uncertainties of
these predictions is given as well.  We also present numerical
results for the $H \to J/\psi \,\gamma$ and $H \to \Upsilon(1S)\,\gamma$
branching ratios in the SM and study the impact of deviations of
the $H\bar{c}c$ and $H\bar{b}b$ couplings from their SM values.

\subsection{Direct-production amplitude}

We begin with the direct production mechanism.  This mechanism was first
discussed in Ref.~\cite{Keung:1983ac}.  We will allow the
heavy-quark Yukawa couplings to deviate from their SM values by factors
$\kappa_Q$:
\begin{equation}
g_{H\bar{Q}Q} = \kappa_Q \left(g_{H\bar{Q}Q}\right)_{\rm SM},
\label{eq:scale}
\end{equation}
where $Q=c,b$.
The amplitude for the direct production process can be computed in the
nonrelativistic QCD (NRQCD) factorization framework \cite{Bodwin:1994jh}
as an expansion in powers of $\alpha_s$ and $v$, where $v$ is the
heavy-quark or heavy-antiquark velocity in the quarkonium rest frame.
The result at leading order in $\alpha_s$ and $v$ is
\begin{equation}
{\cal M}_{direct} = \frac{4\sqrt{3} e_Q e \kappa_Q}{m_H^2-m_V^2} \left[
\frac{G_F m_V }{2\sqrt{2}}\right]^{1/2}\phi_0(V)
	\left\{2p_\gamma \cdot \epsilon_V^{*} p_V \cdot
\epsilon_\gamma^{*}-(m_H^2-m_V^2) \epsilon_\gamma^{*} \cdot \epsilon_V^{*}  
\right\},
\label{eq:direct}
\end{equation}
in agreement with Ref.~\cite{Keung:1983ac}.  Here, $e_Q$ is the
heavy-quark electric charge in units of $e$, $m_Q$ is the
heavy-quark mass, $G_F$ is the Fermi constant, and $\epsilon_\gamma$
and $\epsilon_V$ denote the polarization vectors of the photon and
vector meson, respectively. The factor, $\phi_0$ is the wave
function of the quarkonium state at the origin. Numerical values for
this factor are
\begin{eqnarray}
\phi_0^2(J/\psi)&=&0.073_{-.009}^{+0.011}~\text{GeV}^3 
\qquad\hbox{(Ref.~\cite{Bodwin:2007fz})},\nonumber\\
\phi_0^2(\Upsilon)&=&0.512_{-.032}^{+.035}~\text{GeV}^3
\phantom{3}\qquad\hbox{(Ref.~\cite{Chung:2010vz})}.
\label{phi_0-values}
\end{eqnarray}
We note that the next-to-leading-order (NLO) QCD correction to this
process has been calculated previously~\cite{Vysotsky:1980cz} and
decreases the direct-production rate by approximately 50\%.  The
large size of this correction is due to large logarithms of the form
$\text{ln} (m_H^2/m_V^2)$, which must be resummed in order to obtain a
reliable perturbative expansion. There are two sources of such
logarithms: the emission of collinear gluons and the running of the
$H\bar{Q}Q$ coupling. These logarithms have been resummed in the
leading-logarithmic approximation~\cite{Shifman:1980dk}, and the
resummed result, including the full NLO correction, is 
smaller than the Born amplitude for direct production by a factor of $0.597$ 
for the $J/\psi\,\gamma$ final state and by a factor of $0.689$ for the 
$\Upsilon(1S)\,\gamma$ final state. We include this
resummed QCD correction in the numerical results that we
present.

\subsection{Indirect-production amplitude}

Next we calculate the amplitude for indirect production through a
virtual photon. Here,  following the treatment in
Ref.~\cite{Bodwin:2006yd}, we note that the virtual photon couples to
a vector quarkonium through a matrix element of the 
electromagnetic current. The scattering amplitude that corresponds to this
matrix element of the electromagnetic current is 
\begin{equation}
i{\cal M}_{J_V}=-ie\langle V(\epsilon)|J_V^\mu(x=0)|0\rangle =-ieg_{V\gamma}
\epsilon^{\mu*},
\label{vector-EM-ME}
\end{equation}
where $\epsilon$ is the polarization vector of the quarkonium and $J_V$
is the electromagnetic current: 
\begin{equation}
J_V^\mu(x)= \sum_q e_q \bar q(x)\gamma^\mu q(x).
\label{em-current}
\end{equation}
In Eq.~(\ref{em-current}), the sum is over all quark flavors, heavy
and light. The decay of the quarkonium $V$ to a lepton-antilepton pair
through a single virtual photon is mediated by the adjoint of the matrix
element in Eq.~(\ref{vector-EM-ME}). It follows straightforwardly that
\begin{equation}
\Gamma[V\to l^+l^-]=\frac{4\pi\alpha^2 g_{V\gamma}^2}{3m_V^3}.
\label{leptonic-width}
\end{equation}
Therefore, the magnitude of the coupling $g_{V\gamma}$
can be determined from Eq.~(\ref{leptonic-width}). In order to 
determine the phase of $g_{V\gamma}$, we note that the matrix element of 
the electromagnetic current in a vector quarkonium state
has an expansion in terms of NRQCD operator matrix elements:
\begin{equation}
\langle V(\epsilon)|J_V^\mu(x=0)|0\rangle=
\sum_n c_n\langle V|{\cal O}_n|0\rangle,
\end{equation}
where the short-distance coefficients $c_n$ have an expansion in powers
of $\alpha_s$, and the long-distance matrix elements of the NRQCD
operators ${\cal O}_n$ scale as known powers of the heavy-quark velocity
$v$. At leading order in $\alpha_s$ and $v$ we have, in the quarkonium
rest frame,
\begin{equation}
\langle V(\epsilon)|J_V^\mu(x=0)|0\rangle=g^{\mu}_i e_Q
\langle V(\epsilon)|{\cal O}^i(^3S_1^{[1]}|0\rangle
=- \sqrt{2N_C}\sqrt{2m_V}\phi_0 e_Q \epsilon^{*\mu},
\end{equation}
from which it follows that 
\begin{equation}
g_{V\gamma}=-e_Q\sqrt{2N_c}\sqrt{2m_V}\, \phi_0.
\label{gvgam-phi0}
\end{equation}
We take $\phi_0$ to be real. An imaginary contribution to $g_V$ of
higher order in $\alpha_s$ arises from the annihilations of light
quark-antiquark pairs into the heavy quark-antiquark pair of the
quarkonium. This contribution affects the short-distance coefficient of
the NRQCD matrix element $\langle V| \chi^\dagger {\bm
\sigma}\psi|0\rangle$ and is of relative order $\alpha_s^3(m_V)$. An
imaginary contribution to $g_V$ of higher order in $v$ arises from the
production of  of the $Q\bar Q gg$ Fock state of the $J/\psi$. This
contribution affects the NRQCD matrix element $\langle V|\chi^\dagger
{\bm D}^2{\bm\sigma}\psi|0\rangle$ and is of relative order $v^6$.
Therefore, Eq.~(\ref{gvgam-phi0}) determines the phase of $g_{V\gamma}$
relative to the phase of $\phi_0$, up to corrections that are negligible
in comparison to other uncertainties in the calculation.

The method of computation that is based on Eq.~(\ref{vector-EM-ME}) has
an important advantage over a straightforward calculation in the
framework of NRQCD factorization
\cite{Bodwin:1994jh}. A calculation in the NRQCD factorization framework
proceeds through a double expansion in powers of $\alpha_s$ and $v$. The
NRQCD long-distance matrix elements that enter at each order in $v$ are
nonperturbative quantities that must be determined from lattice
calculations or from phenomenology. On the other hand, the matrix element of
the electromagnetic current in Eq.~(\ref{vector-EM-ME}) already contains
all of the corrections of higher order in $\alpha_s$ and $v$ that appear
in either quarkonium decay through a single virtual photon or quarkonium
production through a single virtual photon. Hence, the higher-order
corrections in both $\alpha_s$ and $v$ cancel when one uses
Eq.~(\ref{leptonic-width}) to express the production amplitude in terms
of the decay width.

From Eq.~(\ref{vector-EM-ME}), we find that the indirect-production 
amplitude is given by 
\begin{equation}
{\cal M}_{indirect}=-e \frac{g_{V\gamma}}{m_V^2}{\cal M}_{H\to 
\gamma\gamma},
\label{indirect-amp}
\end{equation}
where we have used 
\begin{equation}
{\cal M}_{H\to \gamma\gamma^*}= {\cal M}_{H\to \gamma\gamma}+O(m_V^2/m_H^2). 
\label{gamma*-gamma}
\end{equation}
From existing calculations of $ {\cal M}_{H\to \gamma\gamma}$, it 
follows that
\begin{equation}
{\cal M}_{indirect} = -e \frac{\alpha}{\pi} \frac{g_{V\gamma}}{m_V^2}\left( \sqrt{2}G_F\right)^{1/2} {\cal I} \left[2p_\gamma \cdot \epsilon_V^{*} p_V \cdot \epsilon_\gamma^{*}-(m_H^2-m_V^2) \epsilon_\gamma^{*} \cdot \epsilon_V^{*}  \right].
\label{eq:indirect}
\end{equation}
${\cal I}$ denotes the loop-induced coupling of the Higgs to photons,
which arises primarily from top-quark and $W$-boson loops.  Its value at 
leading order in $\alpha_s$ can be
found in Ref.~\cite{Spira:1995rr}.  It is known through NLO in $\alpha_s$ 
(Refs.~\cite{Spira:1995rr,Zheng:1990qa,Djouadi:1990aj,Dawson:1992cy,Melnikov:1993tj,Inoue:1994jq}).  The two-loop electroweak corrections to this quantity are also known~\cite{Actis:2008ts}.  We
can combine the amplitudes in Eqs.~(\ref{eq:direct})
and~(\ref{eq:indirect}) to obtain the following decay width:
\begin{equation}
\Gamma(H \to V\gamma) = \frac{1}{8\pi} \frac{m_H^2-m_V^2}{m_H^2}
|{\cal A}_{\rm direct}+{\cal A}_{\rm indirect}|^2,
\end{equation}
where
\begin{subequations}
\begin{eqnarray}
{\cal A}_{\rm direct}&=&2\sqrt{3} e_Q e \kappa_c
\left(\sqrt{2}G_F m_V\right)^{1/2}
\frac{m_H^2-m_V^2}{\sqrt{m_H}(m_H^2-m_V^2/2-2m_Q^2)} 
\phi_0(V),\\
{\cal A}_{\rm indirect}&=&
-\frac{eg_{V\gamma}}{m_V^2} (\sqrt{2}G_F)^{1/2} \frac{\alpha}{\pi} 
\frac{(m_H^2-m_V^2)}{\sqrt{2 m_H}} {\cal I}.
\end{eqnarray}
\end{subequations}
We have included the $m_V^2$ dependence of ${\cal A}_{\rm indirect}$
that arises from the tensor in Eq.~(\ref{eq:indirect}) and the exact
$m_V^2$ dependences of the phase space and of ${\cal A}_{\rm direct}$.
We note that ${\cal I}$ is negative for relevant values of the Higgs
mass, except for a small phase of about $0.005$. Using 
Eq.~(\ref{gvgam-phi0}), where it can be seen that $g_{V\gamma}$ contains 
a factor $e_Q$, we find that the interference between production
mechanisms is destructive for both the $J /\psi$ and the $\Upsilon(1S)$
final states.  Making use of Eq.~(\ref{indirect-amp}), we can write the
indirect contribution to ${\cal A}$ in terms of the $H \to \gamma
\gamma$ amplitude:
\begin{equation}
{\cal A}_{\rm indirect} = \frac{eg_{V\gamma}}{m_V^2} 
\left[16\pi\Gamma(H\to\gamma\gamma)\right]^{1/2}
\frac{m_H^2-m_V^2}{m_H^2}
\left[1-\left(\frac{m_V}{183.43~{\rm GeV}}\right)^2\right],
\label{A-indir-alternate}
\end{equation}
where we have multiplied $\Gamma(H\to\gamma\gamma)$ by a factor of $2$ to
remove its identical-particle symmetrization factor, and we have dropped
the small phase of ${\cal I}$, which affects the interference term in
the cross section by an amount that is completely negligible in
comparison with the theoretical uncertainties in the interference term.
In Eq.~(\ref{A-indir-alternate}), we have included the corrections of
order $m_V^2/m_H^2$ to $\Gamma(H\to\gamma\gamma^*)$ that appear at
leading order in $\alpha_s$. These can be inferred from the results for
the $HZ\gamma$ coupling that are given in Ref.~\cite{Djouadi:1996yq}.
In our numerical analysis, we obtain ${\cal A}_{H\to
\gamma\gamma}$ from the results for the $H\to \gamma\gamma$ branching
ratio and the $H$ total width in
Refs.~\cite{Dittmaier:2011ti,Dittmaier:2012vm}.
This has the effect of incorporating
higher-order radiative corrections into our prediction.

\subsection{Numerical results}

The contribution of the indirect-production amplitude to
the $H \to V\gamma$ rate can be calculated in the SM with a precision of
a few percent.\footnote{Uncertainties in the indirect widths arise as
follows. The leading correction to the single-virtual-photon
quarkonium-production amplitude arises from triple-gluon quarkonium
production, where one gluon has energy of order $m_V$ and the other two
gluons have energies of order $m_V v$ in the quarkonium rest frame. This
correction is suppressed as $\alpha_s^{3/2}(m_t)
\alpha_s^{1/2}(m_V)v^2(m_V^2/m_t^2)/(\pi \alpha)$ relative to the
amplitude that we compute. The suppression factor is about $7\times
10^{-5}$ for the $J/\psi$ and $3\times 10^{-4}$ for the
$\Upsilon(1S)$.  The
theoretical uncertainty from uncalculated higher-order corrections to
$\Gamma(H\to \gamma\gamma)$ is estimated to be 1\%
(Ref.~\cite{Dittmaier:2012vm}). The uncertainties in $m_t$ and $m_W$
result in uncertainties in $\Gamma(H\to \gamma\gamma)$ of about
$2.2\times 10^{-4}$ and $2.4\times 10^{-4}$, respectively. The
uncertainties in $g_V^2$ that arise from the uncertainties in the
quarkonium leptonic widths are about $2.5\%$ for the $J/\psi$ and about
$1.3\%$ for the $\Upsilon(1S)$. Adding these uncertainties in quadrature,
we conclude that the uncertainty in $\Gamma_{\rm indirect}(H\to J/\psi
\gamma)$ is about $2.7\%$ and the uncertainty in $\Gamma_{\rm
indirect}(H\to \Upsilon(1S)\, \gamma)$ is about $1.6\%$. We have not
included the uncertainty in $\Gamma_{\rm indirect}(H\to V\gamma)$ that
arises from the uncertainty in $m_H$. For a 1~GeV uncertainty in $m_H$,
this is an uncertainty of about $3.5\%$. However, if $m_H$ is ultimately
measured with a precision of about $0.1\%$, then this source of
uncertainty will become negligible.} Hence, these uncertainties should
not be an obstacle in discerning the contribution that arises from the
direct production mechanism. Our predictions for the indirect
contributions for $m_H=125$~GeV are
\begin{subequations}
\begin{eqnarray}
\Gamma_{\rm indirect}(H\to J/\psi\, \gamma)&=& (1.32\pm 0.04)
\times 10^{-8}~{\rm GeV},\\
\Gamma_{\rm indirect}(H\to \Upsilon(1S)\, \gamma)&=& (1.02\pm 0.02)
\times 10^{-9}~{\rm  GeV},
\end{eqnarray}
\end{subequations}
where we have used $\alpha(m_{J/\psi})=1/132.64$ and 
$\alpha(m_{\Upsilon(1S)})=1/131.87$ in computing the photon-quarkonium 
couplings.

The principal uncertainties in the direct-production amplitudes arise
from $\phi_0$ [Eq.~(\ref{phi_0-values})], from uncalculated
corrections of order $\alpha_s^2$, which are not included in the
calculation of Ref.~\cite{Shifman:1980dk}, and from uncalculated
corrections of order $v^2$. We estimate the order-$\alpha_s^2$ corrections
to be $2\%$ and the order-$v^2$ corrections to be
$30\%$ for the $J/\psi$ and $10\%$ for the $\Upsilon(1S)$. We make use
of these uncertainty estimates to obtain the following predictions for
the SM widths of $H$ into quarkonium plus photon for $m_H=125$~GeV:
\begin{subequations}
\begin{eqnarray}
\Gamma_{\rm SM}(H\to J/\psi\, \gamma)&=& (1.00_{-0.10}^{+0.10})
\times 10^{-8}~{\rm GeV},\\
\Gamma_{\rm SM}(H\to \Upsilon(1S)\, \gamma)&=& (5.74_{-4.64}^{+8.27})
\times 10^{-11}~{\rm  GeV}.
\end{eqnarray}
\end{subequations}
In computing the direct amplitudes, we have used
$\alpha=\alpha(m_H/2)=1/128$. In order to maintain compatibility with
the result of Ref.~\cite{Shifman:1980dk}, which is given in terms of the
heavy-quark pole mass, we have set $m_c=m_c({\rm pole})=(1.67\pm
0.07)~{\rm GeV}$ and $m_b=m_b({\rm pole})=(4.78\pm 0.06)~{\rm GeV}$ in
the direct amplitudes. We note that the $H\to J/\psi \, \gamma$ rate
is under reasonably good theoretical control.

In order to get a feeling for the sizes of the SM rates that are
associated with these production modes, we convert them to branching
ratios, using the result for the total Higgs width that is given in
Ref.~\cite{Dittmaier:2012vm}. We obtain the following results for
$J/\psi$ and $\Upsilon$ decays:
\begin{subequations}\label{BRs}
\begin{eqnarray}
\text{BR}_{\rm SM}(H \to J/\psi\,\gamma) &=& (2.46_{-0.25}^{+0.26}) 
\times 10^{-6}, \\
\text{BR}_{\rm SM}(H \to \Upsilon(1S)\,\gamma) &=& (1.41_{-1.14}^{+2.03}) 
\times 10^{-8}.
\end{eqnarray}
\end{subequations}
We note that, for the $J/\psi$ final state, consideration of the
direct amplitude alone would lead instead to a branching ratio of $5.48 \times
10^{-8}$, while for the $\Upsilon(1S)$ it would lead to $3.84
\times 10^{-7}$.   The inclusion of the indirect amplitude is crucial in 
order to obtain an
accurate prediction for the $V\gamma$ production rate.  In order to
compute the rate for the experimentally clean $l^+l^-$ final state, we 
must multiply these results
by branching ratios for $V \to l^+ l^-$,
which are 5.93\% for the $J/\psi$ and 2.48\% for the $\Upsilon(1S)$, with
$l=e$ or $\mu$. We will estimate the event yields more carefully in
Sec.~\ref{sec:exp}, but for now we simply multiply the branching
ratios above by the inclusive cross sections that are tabulated
in Ref.~\cite{LHCXS} in order to determine the number of events
that will be produced at the LHC.  It is clear that the $\Upsilon$
event yield in the SM is far too small to be observed experimentally, and so 
we focus on the
$J/\psi$.  Summing over both electron and muon final states and
combining the event yields of ATLAS and CMS, we find 0.3 $J/\psi \to
l^+ l^-$ events for an integrated luminosity of $20 \, {\rm fb}^{-1}$
at an 8~TeV LHC.  This event yield is too small to be observed.
However, an integrated luminosity of $3000 \, {\rm
fb}^{-1}$ at a 14 TeV LHC would produce 100 $J/\psi \to l^+ l^-$
events.  The $J/\psi$ mode should be observable at the high-luminosity
LHC run, as we discuss in more detail in the next section.

As we have mentioned, the quantities that appear in the indirect-production 
amplitude are very well known, and the key quantity that
appears in this amplitude, $\Gamma(H\gamma\gamma)$, will be measured
with increasing precision at the LHC. Therefore, it is possible, in
principle, to distinguish the effect of the amplitude that arises from
direct $H\bar QQ$ coupling from the effect of the indirect-production
amplitude. We note that turning off the direct-production amplitude for
the $J/\psi$ would lead to a branching ratio of $3.25 \times 10^{-6}$
and 132 events.  This is a statistically significant deviation of 
about 30\% from the SM 
event yield. Hence, measurement of the $H\bar c c$ coupling is a reasonable 
goal for
future experimental searches.

\begin{figure}[t]
\centerline{
\includegraphics[height=6.5cm,angle=0]{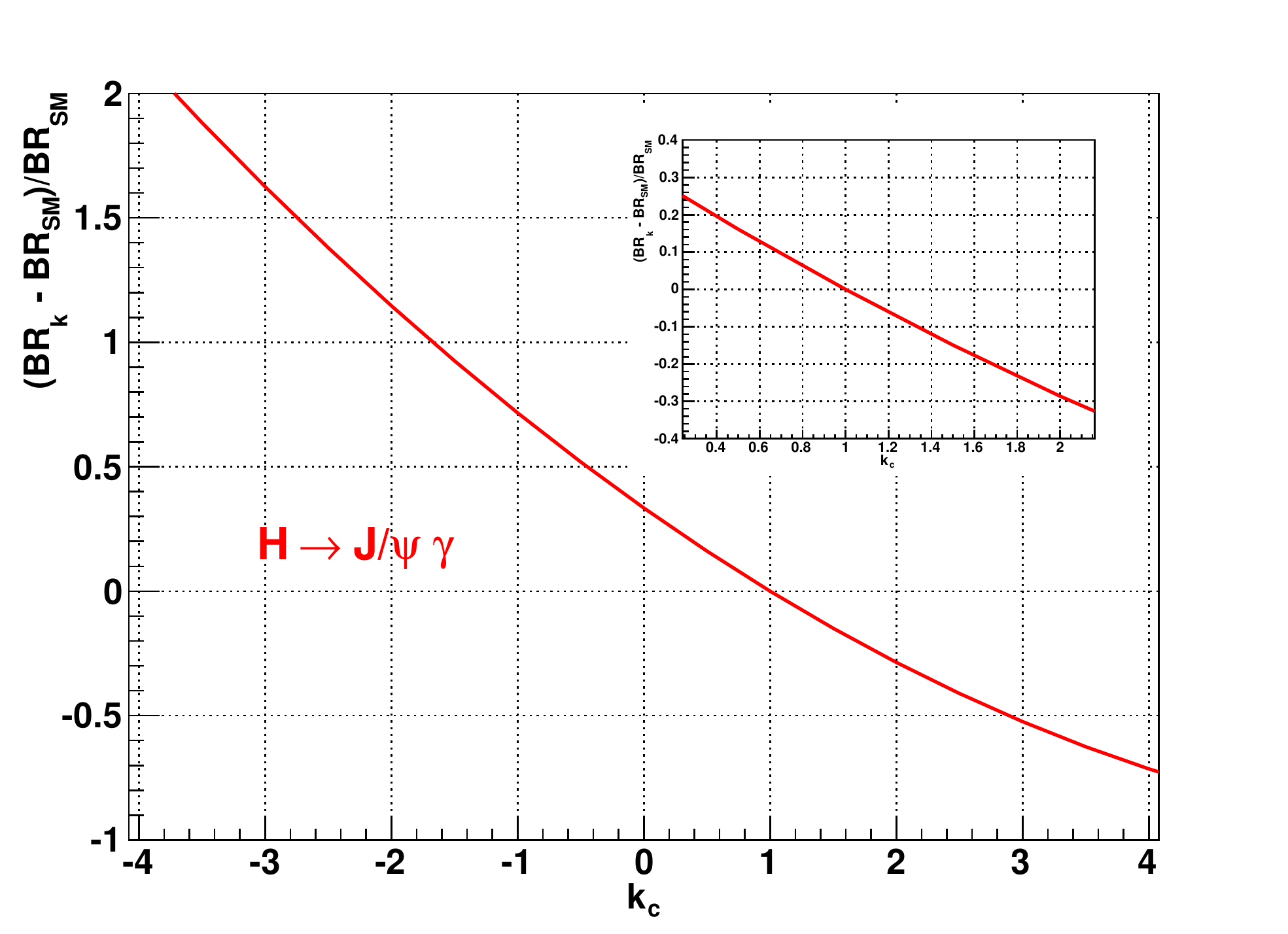}
\hspace{-1.0cm}
\includegraphics[height=6.5cm,angle=0]{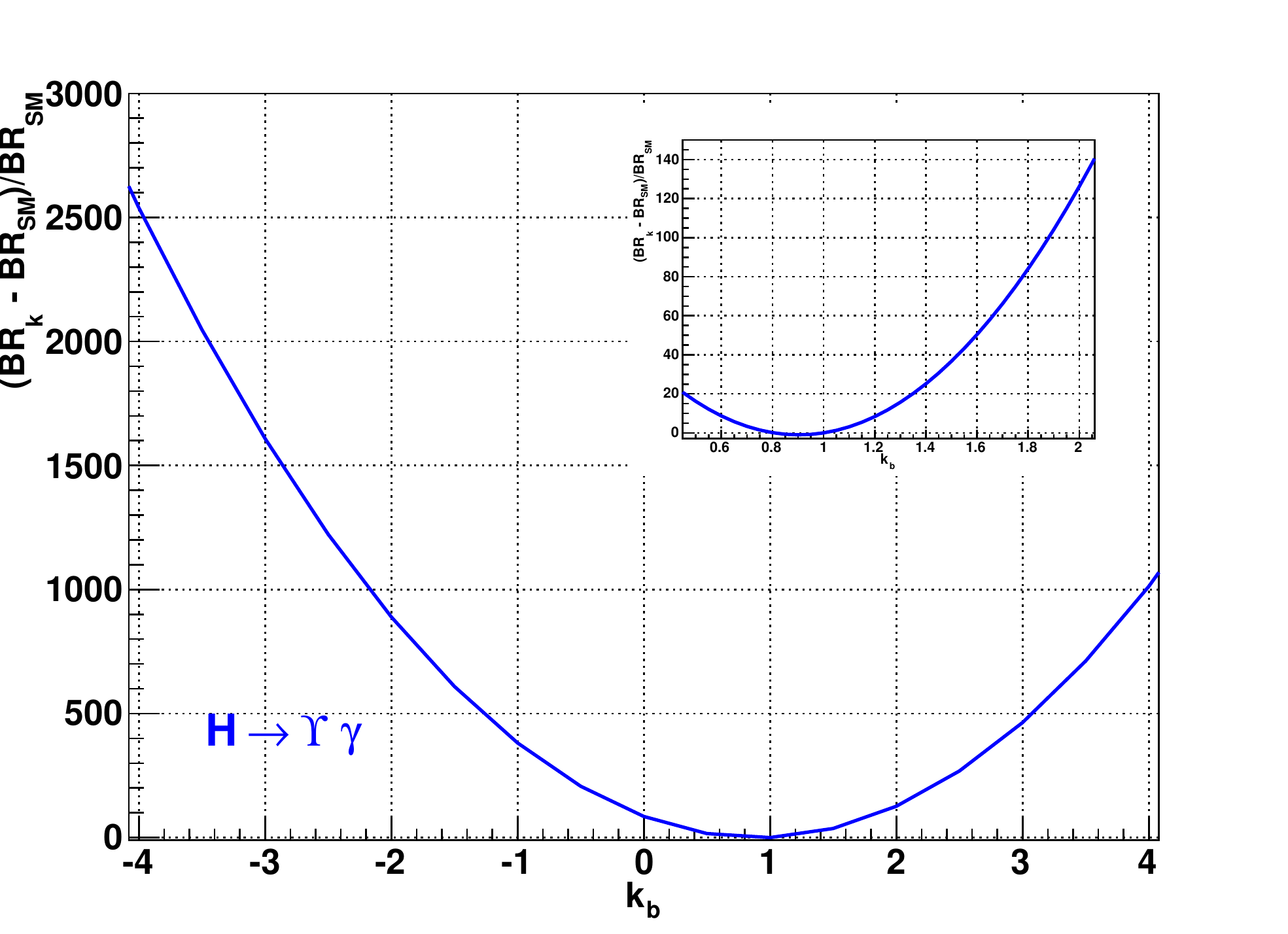}
}
\vspace{-0.5cm}
\caption{The relative deviations in the branching ratios for $H \to
J/\psi \,\gamma$ (left panel) and $H \to \Upsilon(1S) \,\gamma$ (right
panel) as functions of the scaling parameters $\kappa_Q$, which are
defined in Eq.~(\ref{eq:scale}). }
\label{BRdev}
\end{figure}

Deviations of $\kappa_Q$ from unity parametrize deviations of the
$H\bar QQ$ coupling from its SM value. We show in Fig.~\ref{BRdev} the
relative deviations in the $H \to J/\psi \,\gamma$ and $H \to
\Upsilon(1S) \, \gamma$ branching ratios as functions of
$\kappa_Q$.  The shifts in the experimentally promising $J/\psi$ mode
can reach 100\% for values of $\kappa_c$ that are a few times the SM
value.  In the case of $\Upsilon(1S)$ production, the deviations are
extraordinarily large: Within the SM there is a strong cancellation
between the direct and indirect production mechanisms that is lifted if
the $H\bar{b}b$ coupling is changed. Changes in this coupling of a few
times the SM value can, therefore, likely be probed in this channel at
the LHC. Because the interference of the $\Upsilon(1S)$ SM production
amplitudes is almost completely destructive,  most values of $\kappa_b
\ne 1$ result in an increase in the predicted branching ratio relative
to its SM value.

Now let us investigate whether the $J/\psi\gamma$ decay mode is
visible over the continuum $H \to \mu^+\mu^-\gamma$ decay mode. We
estimate the continuum background by integrating the continuum
production rate~\cite{Firan:2007tp} over the range $m_{\mu^+\mu^-} \in
[m_{J\psi}-0.05 \, \text{GeV},m_{J\psi}+0.05 \, \text{GeV}]$. The
integration range is consistent with the experimental resolution, which
is discussed in the next section. We find that
\begin{equation}
\text{BR}_{\rm cont}(H \to \mu^+\mu^-\gamma) = 2.3 \times 10^{-7},
\end{equation}
which is comparable in size to $\text{BR}_{\rm SM}(H \to J/\psi \,
\gamma)\text{BR}(J/\psi\to \mu^+\mu^-)$. Our conclusion is that the 
$J/\psi\gamma$ mode should be visible over the continuum background.

\section{Experimental perspectives}
\label{sec:exp}

The ATLAS and CMS collaborations can search for the $V\gamma$ decay
channels by using the single-lepton, di-lepton or lepton-plus-photon
triggers. The Higgs-to-$V \gamma$ decay is characterized by a
high-$p_{\rm T}$ photon
recoiling against  a lepton-antilepton pair from the  $V$ decay.  The 
vector quarkonium state will be highly boosted, causing the two leptons
to be close to each other in angle, with their  momenta transverse to
the boost axis anti-correlated.  On the basis of these event
characteristics and the current performance of the ATLAS and CMS
detectors and event reconstruction, the following conclusions can be
drawn.
\begin{enumerate}
\item  The resolution of the invariant mass of the lepton and
antilepton is almost independent of their kinematics. The average
lepton momentum is expected to be around 30\,GeV.  Therefore, the 
resolution of the muon transverse momenta ($\mu^+\mu^-$ invariant
mass) can be as good as  $1.3\%$  ($1.8\%$)~\cite{muonres}.
\item The resolution of the photon energy is around
$1\%$~\cite{photonres}.
\item  The  resulting resolution of the three-body (Higgs) invariant mass 
is around 2.1\%. However, if the leptons and the photon are both at high
pseudorapidity, then the resolution will be only about 4\%.
\item The production vertex is well defined by the leptons and, owing
to the high energy of the photon, the contamination from pile-up
events (those with multiple interactions per bunch crossing) is
expected to be small.
\end{enumerate}
\begin{figure}[t]
\centerline{
\includegraphics[height=6.5cm,angle=0]{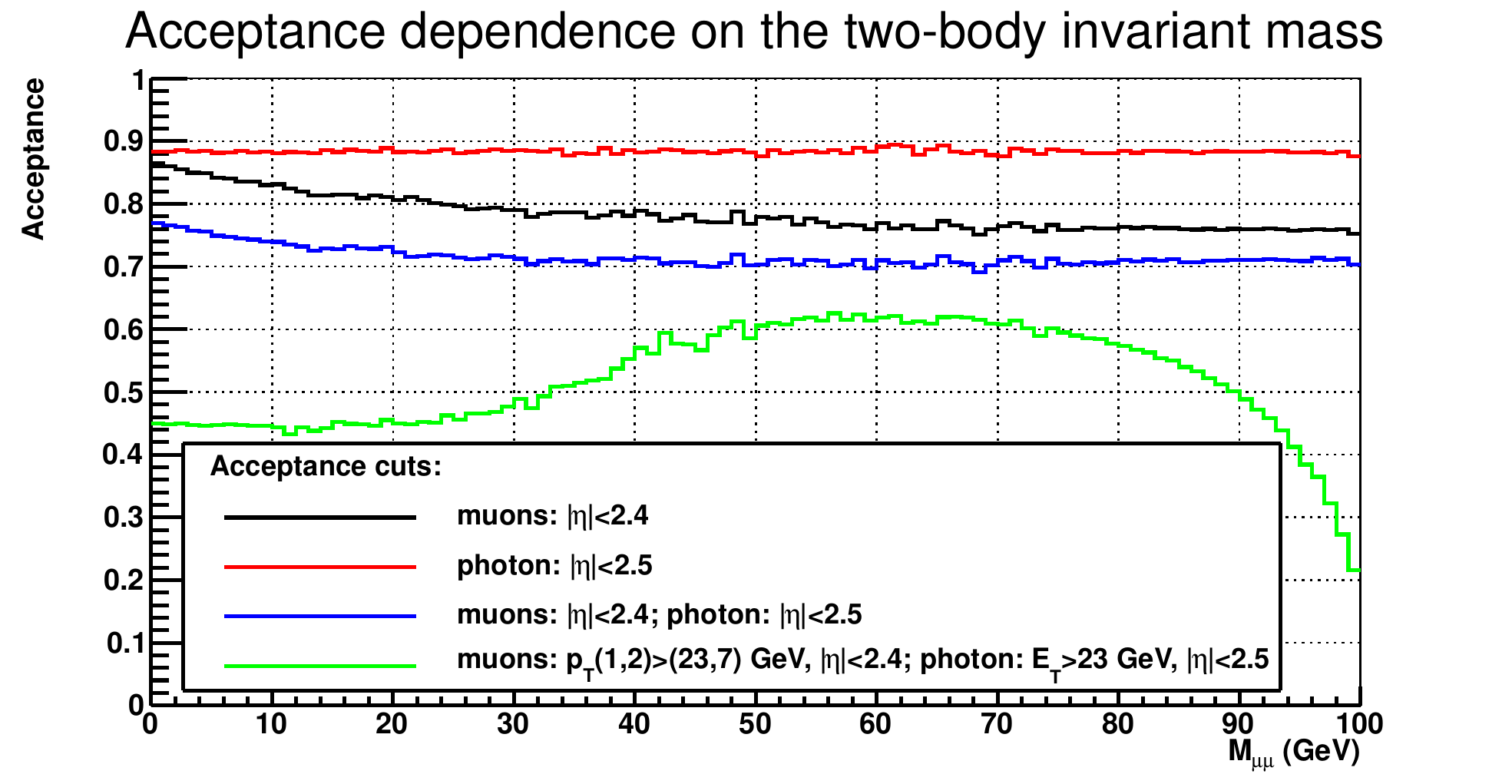}
}
\vspace{-0.3cm}
\caption{Top curve: Geometrical acceptance for the Higgs-to-$\mu\mu\gamma$ 
decay channel for $m_H = 125$\,GeV. Lower  curves: 
Acceptance after the application of additional kinematic requirements.}
\label{acc}
\end{figure}
\begin{figure}[b]
\centerline{
\includegraphics[height=7.5cm,angle=0]{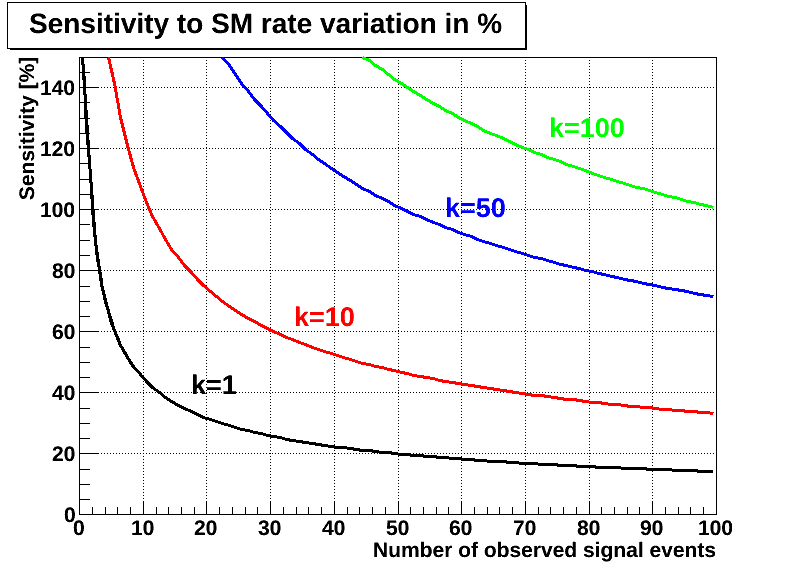}
}
\vspace{-0.3cm}
\caption{Search sensitivity for the process  $H \to J/\psi \, \gamma 
\to l^+l^-\gamma$ as a function of the number of expected
signal events. $k$ is the ratio of background over signal; as discussed in 
the text, $k<10$ is expected in the 
experimental analysis. If one combines the events in the muon and
electron channels and combines the ATLAS and CMS data, then about 50
reconstructed  $H \to J/\psi \, \gamma \to l^+l^-\gamma$ signal events
are expected for an integrated luminosity of 3000\,fb$^{-1}$ and at a
center-of-mass energy of 14~TeV.}
\label{sensitivity}
\end{figure}
As is shown in Fig.~\ref{acc}, studies that are based on the
MCFM~\cite{Campbell:2010ff} event generator predict that the detector
geometrical acceptance for Higgs-to-$\mu\mu\gamma$ events is better than
70\%. After a basic event selection has been performed,  45--60\% of the
signal events will remain. Since there is no missing energy in the
signal events and the expected mass resolution is a few GeV, a clear
resonance over the background in the $\mu\mu\gamma$  invariant mass
distribution is expected. To first approximation,  the sensitivity of
the measurement is given by $\sqrt{(S+B)}/S$, where $S$ and $B$ are the
signal and background events, respectively. The numerator corresponds to
the statistical uncertainty of the observed sample.
Figure~\ref{sensitivity} shows  the expected  sensitivity  as a function
of the signal events at different values of $k=B/S$ (background-to-signal 
ratio). On the basis of the $H\to Z\gamma$ search at ATLAS and
CMS~\cite{Zgam}, we expect the performance and sensitivity of the ATLAS and CMS
detectors for $H\to J/\psi\gamma$ in the electron channel to be similar
to that for $H\to J/\psi\gamma$ in the muon channel.

Given the sensitivity that is required to observe the process $H\to
\gamma\gamma$ at the LHC, we estimate that a sensitivity of about
30--40\% is required in order to observe the process $H \to V\gamma$ at
the LHC. The current $H\to \gamma \gamma$ searches,which were
performed using the 8\,TeV data, observed about 400 signal events per
experiment in a mass window around 125~GeV, with a
background-to-signal ratio that is estimated to be about 50. In an
$H \to J/\psi\,\gamma$ search, the background-to-signal ratio is
expected to be 10 or lower after one has imposed the requirement
that the di-lepton pair and the photon be back-to-back and the
requirement that the di-lepton invariant mass be consistent with
the  $J/\psi$ mass. Suppose that an overall acceptance and
event-reconstruction efficiency of 50\% is achieved and that one
combines the events in the electron and the muon decay channels and
combines the ATLAS and CMS data. Then, 50 signal events can be expected
for an integrated luminosity of 3000\,fb$^{-1}$ at a center-of-mass
energy of 14~TeV. As is shown in  Fig.~\ref{sensitivity}, this data
sample could be large enough for one to observe the $H\to
J/\psi\,\gamma$ decay channel at the LHC.  If a background-to-signal
ratio of unity can be achieved, then the measurement may be sensitive
to the direct-production amplitude, and, therefore, to the
$H\bar{c}c$ coupling in the SM.

\section{Conclusions}
\label{sec:conc}

In this paper we have reconsidered the decays $H \to V \gamma$, with
$V =J/\psi, \Upsilon (1S)$.  We have identified a previously unstudied
mechanism for this decay: $H \to \gamma^{*}\gamma$, followed by the
transition $\gamma^{*} \to V$.  This indirect production mechanism
is the dominant contribution to quarkonium production in Higgs decays,
and leads to a production rate for the $J /\psi \gamma$ final state
that is an order of magnitude larger than had previously
been estimated.

The indirect production mechanism interferes at the amplitude level with
the direct production mechanism, which proceeds via the $H\bar{Q}Q$
coupling.  This interference enhances the effect of the
direct-production amplitude on the $H\to V\,\gamma$ decay rate, opening
the possibility that the $H\bar{Q}Q$ coupling can be measured at the
LHC.  In the SM, the interference term shifts the $H \to J/\psi \,
\gamma$ rate by 30\%.  If the $H\bar{c}c$ coupling deviates from its SM
value by a factor of two or more, then this shift can reach 100\% or
more. In the case of the $H\to \Upsilon(1S) \,\gamma$ decay rate, for
which there is an almost complete cancellation between the direct and
indirect amplitudes in the SM, a deviation of the $H\bar{b}b$ coupling
from its SM value by a factor of two or more can shift the decay rate by
a factor of 1000 or more. We have argued that the indirect-production
amplitude is known with few-percent accuracy within the SM. Therefore,
the uncertainty in indirect-production amplitude would not preclude the
measurement of an $H\bar{c}c$ coupling that is of order the SM value or
an $H\bar{b}b$ coupling that is a few times the SM value.

We have presented numerical results for both the $J/\psi\,\gamma$ and
$\Upsilon (1S)\,\gamma$ final states, and we have performed a
realistic analysis of the $J/\psi \to l^+l^-$ signal at the LHC.  At
a high-luminosity LHC that has accumulated several inverse attobarns of integrated
luminosity, the $l^+l^-$ decay mode should be observable. 
Consequently, it may be possible to detect the effect of the
direct-production amplitude, and thereby to obtain a direct
measurement of the $H\bar{c}c$ coupling at the LHC.

We conclude that the $J/\psi\,\gamma$ decay mode of the Higgs may enable
the direct measurement of the $H\bar{c}c$ coupling at the LHC---something
that was previously believed to be impossible. Such a measurement would
provide a further test of the hypothesis that the observed
Higgs-like particle has the couplings of an elementary SM Higgs. We
believe that the possibility of observing the $H\bar{c}c$ coupling
through the $J/\psi\,\gamma$ decay mode provides a motivation for the
high-luminosity run of the LHC, and we encourage the ATLAS and CMS
collaborations to pursue this measurement.

\begin{acknowledgements}
We thank Heather Logan for a helpful discussion regarding the use of
Higgs decays to quarkonia to resolve sign ambiguities in Higgs-coupling
determinations.  The work of G.B. is supported by the U.S.\ Department of Energy, Division
of High Energy Physics, under contract DE-AC02-06CH11357.  The work of F.P. is supported by the U.S.\ Department of Energy, Division
of High Energy Physics, under contract DE-AC02-06CH11357 and the grant DE-FG02-91ER40684. The work of S.S. and M.V. is supported by the U.S.\ Department of Energy, Division
of High Energy Physics, under the grant DE-FG02-91ER40684.

The submitted manuscript has been created in part by UChicago Argonne, LLC, Operator of Argonne National Laboratory (ÒArgonneÓ). Argonne, a U.S. Department of Energy Office of Science laboratory, is operated under Contract No. DE-AC02-06CH11357. The U.S. Government retains for itself, and others acting on its behalf, a paid-up nonexclusive, irrevocable worldwide license in said article to reproduce, prepare derivative works, distribute copies to the public, and perform publicly and display publicly, by or on behalf of the Government.

\end{acknowledgements}

\end{document}